\documentstyle[twocolumn,aps,epsfig]{revtex}
\begin{document}
\draft
\title{Net Charge on a Noble Gas Atom Adsorbed on a Metallic Surface}
\author{A. Widom}
\address{Physics Department, Northeastern University, Boston MA 02115}
\author{M.S. Tomassone}
\address{Levitch Institute \& Department of Physics, City College of New 
York, NY NY 10031}
\author{Y.N. Srivastava}
\address{Physics Department \& INFN, University of Perugia, Perugia Italy}
\author{M. Hannout}
\address{Physics Department, University of New Hampshire, Durham NH 03824}
\maketitle

\begin{abstract}
Adsorbed noble gas atoms donate (on the average) a fraction of an 
electronic charge to the substrate metal. The effect has  
been experimentally observed as an adsorptive change in the 
electronic work function. The connection 
between the effective net atomic charge and the binding energy 
of the atom to the metal is theoretically explored. 
\end{abstract}  

\narrowtext

\section{Introduction}

The problem of measuring and understanding the binding energy of noble 
gas atoms on metal substrates has always been of considerable interest 
\cite{1,2,3,4,5,6}. The atomic binding is that energy released when an 
atom from the vapor sticks to the surface. Several studies have related 
atomic binding to surface charge 
distributions\cite{1,7,8,9,10,11,12,13,14,15}. Induced 
surface electronic dipole moments near the adsorbed atom have been 
of particular interest. In the work which follows, the intimate 
relationship between binding energy and induced atomic charge will 
be considered in theoretical detail.

As the atom is lowered onto the surface and becomes adsorbed, the 
atomic dipole tends to be oriented with the positive side of the dipole  
pointing away from the metal. In reality, the negative side of the atomic 
dipole is better thought to be {\em on average} an electron (negative) 
charge \begin{math}-Z_{eff}|e|\end{math} donated to the metal. 
This leaves the atom with a positive net mean charge 
\begin{math}+Z_{eff}|e|\end{math}. The physical situation is 
pictured in FIG.1 below. Even in a situation often regarded as 
physisorption, one does not expect a noble gas atom to remain in 
perfect charge neutrality. When the atom is adsorbed on the substrate, 
the negative end of the atomic dipole moment neutralizes the 
positive end of the {\em image} dipole moment. This leaves a mean net 
positive charge on the atom and a mean negative electronic charge 
deposited in the metal. From an experimental viewpoint, the positive 
nature of the mean charge on an adsorbed noble gas atom is equivalent 
to the donation of a negative charge to the metal. The physical effect 
is made manifest
\cite{1,2,3,9,10,11,12,13,14,16,18} 
by the diminution of the electronic work function as the first monolayer 
of atoms is deposited on the metallic substrate surface. 
This effect is quite large and is observed for all combinations of 
gases and metals\cite{7}. The magnitude of the reduction of the work function 
is proportional to the amount of adsorbate and for coverage up to a 
monolayer\cite{15}.

\begin{figure}[htbp]
\begin{center}
\mbox{\epsfig{file=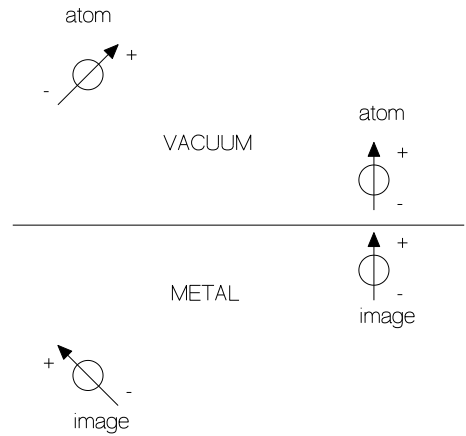,height=80mm}}
\caption{Shown is the surface between the vacuum and the metal. When 
an atom is far above the surface between the vacuum and the metal 
the dipole fluctuations are correlated with the image dipole, as shown 
to the left. As the atom is lowered onto the surface, the atomic dipole 
is pointed upwards as shown on the right. An electron is then donated 
to the metal with probability giving the effective charge strength 
$Z_{eff}$.}
\label{fig1}
\end{center}
\end{figure}

In Secs. II and III  the long ranged part of Van der Waals interaction 
between a noble gas atom and a metal will be reviewed. The height of the 
atom, over and above the metal surface, is considered to be large 
compared with the atomic size. The interaction is between a quantum 
fluctuating dipole and its correlated image. Both the binding energy 
and the probability of electronic excitation will be computed numerically 
in Sec. IV. In Sec. V, non-perturbative expressions for the binding energy 
and the probability of electronic excitation are derived. In Sec. VI it 
is shown that the stronger the binding energy the larger the  
net atomic adsorptive charges. The physical importance of the 
adsorptive charge on a noble gas atom is discussed in the concluding 
Sec. VII.  

\section{Perturbation Theory}

Suppose a Hamiltonian of the  form 
\begin{equation}
{\cal H}=H+V, 
\end{equation}
in which 
\begin{equation}
H\psi_n=E_n\psi_n,
\end{equation}
represents an unperturbed energy eigenvalue problem, and 
\begin{equation}
{\cal H}\Psi_n={\cal E}_n\Psi_n,
\end{equation} 
represents the perturbed energy eigenvalue problem. Without loss of 
generality, one may assume (for the unperturbed ground state wave 
function) that 
\begin{equation}
\left(\psi_0,V\psi_0\right)=0.
\end{equation}

Two quantities of importance are the exact ground state energy ${\cal E}_0$, 
and the probability that the interaction $V$ introduces an excitation in the  
unperturbed quantum states 
\begin{equation}
P=\sum_{n\ne 0}\left|\left(\psi_n,\Psi_0\right)\right|^2
=1-\left|\left(\psi_0,\Psi_0\right)\right|^2.
\end{equation}
In lowest order perturbation theory, one then finds the energy shift
\cite{19}  
\begin{equation}
U=\left({\cal E}_0-E_0\right)=-\sum_{n\ne 0}
{\left|\left(\psi_n,V\psi_0\right)\right|^2\over \left(E_n-E_0\right)}
+...\ ,
\end{equation}
The ground state wave function in first order perturbation theory\cite{20} 
is given by  
\begin{equation}
\Psi_0=\psi_0-\sum_{n\ne 0}
{\left(\psi_n,V\psi_0\right)\psi_n\over \left( E_n-E_0 \right)}
+...\ .
\end{equation}
The first order wave equation can be used to compute the excitation 
probability $P$ to second order in the perturbation potential; it is  
\begin{equation}
P=\sum_{n\ne 0} \left|{\left(\psi_n,V\psi_0\right) 
\over \left(E_n-E_0\right)}\right|^2 +...\ .
\end{equation}
Let us apply these ideas to the Casimir effect, i.e. the attractive force 
on an atom located at height $h$ above a metallic substrate. The 
final energy is well known. However, the excitation probability of the 
atom will also be calculated.

\section{The Van der Waals Force}

Let the unperturbed problem consist of an isolated atom and an isolated 
metal 
\begin{equation}
H=H_{atom}+H_{metal}.
\end{equation}
With $n=(j,a)$ as a double index referring to the atomic state $j$ and the 
metallic substrate state $a$, we may write 
\begin{equation}
\psi_n=\psi^{atom}_j\psi^{metal}_a,
\end{equation}
with an unperturbed energy  
\begin{equation}
E_n=E^{atom}_j+E^{metal}_a .
\end{equation}

Further, let the interaction between the atom and metal be of the dipole 
form  
\begin{equation}
V=-{\bf \mu \cdot E}
\end{equation}
where ${\bf \mu}$ represents the atomic electric dipole moment operator, 
and the electric field {\bf E} is produced by the metal.
The second order perturbation interaction energy between the atom and 
the metal is usually called the Casimir effect and is given by 
\begin{equation}
U_C=
-\sum_{(a,j)\ne (0,0)}{\left|{\bf \mu}_{j0}{\bf \cdot}{\bf E }_{a0}
\right|^2\over \left(\hbar \omega_{j0}+\hbar \omega^\prime_{a0}\right)},
\end{equation}
where 
\begin{equation}
{\bf \mu }_{j0}=\left(\psi^{atom}_j,{\bf \mu}\psi^{atom}_0\right),
\end{equation}
and
\begin{equation}
{\bf E }_{a0}=\left(\psi^{metal}_a,{\bf E}\psi^{metal}_0\right).
\end{equation}
The atomic Bohr frequencies are defined as  
\begin{equation}
\hbar \omega_{j0}=\left(E^{atom}_j-E^{atom}_0 \right).
\end{equation}
and the metallic Bohr frequencies are defined as  
\begin{equation}
\hbar \omega^\prime_{a0}=\left(E^{metal}_a-E^{metal}_0 \right).
\end{equation} 

For an isotropic atom, one may define a ground state dipole moment 
quantum noise spectral function 
\begin{equation}
S(\omega )={1\over 3}\sum_{j\ne 0} \left|\left(\psi^{atom}_j,{\bf \mu }
\psi^{atom}_0\right)\right|^2 \delta \left(\omega -\omega_{j0}\right), 
\end{equation}
Similarly, one may define the electric field quantum zero point 
fluctuations due to the metal  
\begin{equation}
S_{\bf E}(\omega^\prime )
=\sum_{a\ne 0} \left|\left(\psi^{metal}_a,{\bf E }
\psi^{metal}_0\right)\right|^2 
\delta \left(\omega^\prime  -\omega^\prime_{a0}\right). 
\end{equation}
One may now compute the Casimir energy\cite{21,22,23} 
$U_C$ in terms of these spectral functions; i.e.  
\begin{equation}
U_C=-\left({1\over \hbar}\right)\int_0^\infty \int_0^\infty 
\left( {{S(\omega )S_{\bf E}(\omega^\prime }
\over \omega +\omega^\prime }\right) d\omega d\omega^\prime .
\end{equation}
Finally, one may find the spectral functions by employing the zero point 
quantum fluctuation response theorems\cite{24}. 

For example, if the atomic polarization response to an external electric 
field at complex frequency $\zeta $ (with $\Im m(\zeta )>0$), 
\begin{equation}
\delta \left<{\bf \mu }\right>=\alpha (\zeta )\delta {\bf E}_{ext}, 
\end{equation}
defines the ground state atomic polarizability $\alpha (\zeta )$ for 
a spherical atom, then the fluctuation response theorem theorem asserts 
\begin{equation}
S(\omega )=\left({\hbar \over \pi}\right) \Im m\ \alpha(\omega +i0^+).
\end{equation}
The fluctuation spectral function for the electrostatic field 
${\bf E}=-{\bf \nabla }\phi $ produced by the metal is a bit more 
subtle.

If an external charge density at complex frequency $\zeta $ produces an 
electrostatic potential according to the rule 
\begin{equation}
\delta \left<\phi ({\bf r})\right>=
\int {\cal G}({\bf r},{\bf r}^\prime ,\zeta )
\delta \rho_{ext}({\bf r}^\prime)d^3{\bf r}^\prime ,
\end{equation}  
then the spectral function for zero point electrostatic potential 
fluctuations
$$ 
S_\phi ({\bf r},{\bf r}^\prime ,\omega )=
$$
\begin{equation}
\int_{-\infty}^\infty \cos(\omega t)\Re e
\left<0\left|
\Delta \phi ({\bf r},t)\Delta \phi ({\bf r}^\prime ,0)\right|0
\right>\left({dt\over 2\pi }\right)
\end{equation}
obeys the fluctuation response theorem in the form 
\begin{equation}
S_\phi ({\bf r},{\bf r}^\prime ,\omega )=
-\left({\hbar \over \pi }\right)\Im m 
{\cal G}({\bf r},{\bf r}^\prime ,\omega +i0^+ ).
\end{equation}
For the problem at hand, suppose that the metal is located in 
the half-space $z<0$; e.g. the substrate surface is the $z=0$ 
$x$-$y$ plane. If both ${\bf r}$ and ${\bf r}^\prime $ are in the 
vacuum (i.e. $z>0$ and $z^\prime >0$), then the ``method of images'' 
yields the Greens function 
\begin{equation}
{\cal G}({\bf r},{\bf r}^\prime ,\zeta )=
\left({1\over |{\bf r}-{\bf r}^\prime |}\right)
-\left({\eta (\zeta )\over |{\bf r}-{\bf r}_i^\prime |}\right),
\end{equation}
where  
\begin{equation}
{\bf r}^\prime =(x^\prime ,y^\prime ,z^\prime ),
\end{equation}
has a corresponding ``image'' position 
\begin{equation}
{\bf r}_i^\prime = (x^\prime ,y^\prime ,-z^\prime ),
\end{equation}
and 
\begin{equation}
\eta (\zeta )=\left({\varepsilon (\zeta )-1 \over 
\varepsilon (\zeta )+1)}\right).
\end{equation}
The dielectric response function for the metal $\varepsilon (\zeta )$ 
determines the conductivity $\sigma (\zeta )$ via 
\begin{equation}
\varepsilon (\zeta )=1+\left({4\pi i\sigma (\zeta )\over \zeta }\right).
\end{equation}  
Thus 
\begin{equation}
S_\phi ({\bf r},{\bf r}^\prime ,\omega )=
\left({\hbar \over \pi |{\bf r}-{\bf r}_i^\prime |}\right)
\Im m\ \eta (\omega +i0^+).
\end{equation}
With 
\begin{equation}
{\bf R}=(0,0,h)
\end{equation}
denoting the position of the atom at a height $h$ above the substrate 
surface, the electric field zero point fluctuations of the electric 
field at the atom may be computed via 
$$
S_{\bf E}(\omega )=\lim_{{\bf r}\to {\bf R}}\ 
\lim_{{\bf r}^\prime \to {\bf R}}
$$
\begin{equation}
\left(
{\partial ^2\over \partial x\partial x^\prime}+
{\partial ^2\over \partial y\partial y^\prime}+
{\partial ^2\over \partial z\partial z^\prime}
\right)
S_\phi ({\bf r},{\bf r}^\prime ,\omega ).
\end{equation} 
The differentiation is tedious but direct. It yields 
\begin{equation}
S_{\bf E}(\omega )=
\left({\hbar \over 2\pi }\right)
\left({\Im m\ \eta (\omega +i0^+)\over h^3}\right).
\end{equation}

Substituting Eqs.(22) and (34) into Eq.(20) yields the zero-point 
fluctuation Casimir effect potential  
\begin{equation}
U_C=-\left({C_{atom}\over h^3}\right)
\end{equation}
where 
$$
C_{atom}=\left({\hbar \over 2\pi^2}\right)\times 
$$
\begin{equation}
\int_0^\infty \int_0^\infty
\left({\Im m\ \alpha (\omega +i0^+)\Im m\ \eta (\omega^\prime +i0^+)
\over \omega +\omega^\prime }\right)d\omega d\omega ^\prime . 
\end{equation}
The Casimir form of the Van der Waals potential between an atom and 
a conducting surface is well known. The purpose for reviewing the Van 
der Waals result is that now we may also calculate 
the probability that the atom at a height $h$ is in an excited state. 

The excited state probability in Eq.(5), when evaluated to the 
lowest order perturbation theory in Eq.(8), yields the Casimir excited 
state probability $P_C$. In the Casimir perturbation theory, the 
probability for the atom to be in an excited state contains an 
extra energy denominator. The final result is in the simple form 
\begin{equation}
P_C=\left({v_{atom}\over h^3}\right). 
\end{equation}
The parameter \begin{math} v_{atom} \end{math} is a rough 
measure of the effective atomic volume; It is 
$$
v_{atom}=\left({1\over 2\pi^2}\right)\times 
$$
\begin{equation}
\int_0^\infty \int_0^\infty
\left({\Im m\ \alpha (\omega +i0^+)\Im m\ \eta (\omega^\prime +i0^+)
\over (\omega +\omega^\prime )^2}\right)d\omega d\omega ^\prime . 
\end{equation}

The excited state probability in Eq.(37) is a new result. 
The noble gas atom above the metal, is {\em not} electronically 
inert. The atom in the vacuum is in the ground state, As the 
atom is lowered toward the metal surface, the atomic electronic 
state becomes excited with probability $P>0$. For heights $h$ 
such that $h^3>>v_{atom}$ we have $P\approx P_C$ as in Eq.(37).
Numerical results for the probability $P_C$ then follow (below) 
in a similar manner to numerical results for $U_C$ which have 
been previously computed by other workers.

\section{Numerical Evaluations}

For the purpose of numerical evaluations of parameters associated with 
the Van der Waals interaction, we follow Rauber, 
Klein, Cole and Bruch\cite{25}; They choose a single pole approximation 
for both \begin{math} \alpha (\zeta ) \end{math} and 
\begin{math} \eta (\zeta ) \end{math}. The pole in 
\begin{math} \alpha (\zeta ) \end{math} is parameterized by an  
atomic frequency \begin{math} \omega_a \end{math}, while the pole 
in \begin{math} \eta (\zeta ) \end{math} is parameterized by a plasma 
frequency \begin{math} \Omega_s \end{math}. 

In detail, the approximations read 
\begin{equation}
\alpha^{RKCB} (\zeta )\approx 
\left({\omega_a^2 \alpha_0 \over \omega_a^2-\zeta^2 }\right), 
\end{equation}
along with 
\begin{equation}
\eta^{RKCB} (\zeta )\approx 
\left({\Omega_s^2 g_0 \over \Omega_s^2-\zeta^2 }\right). 
\end{equation}

Eqs.(36), (38), (39) and (40) yield   
\begin{equation}
C_{atom}^{RKCB} = {\hbar \over 8} 
\left\{{\alpha_0 g_0\omega_a\Omega_s\over \omega_a+\Omega_s }\right\},
\end{equation}
and 
\begin{equation}
v_{atom}^{RKCB} = {1\over 8} 
\left\{{\alpha_0 g_0\omega_a\Omega_s\over ( \omega_a+\Omega_s )^2}\right\}.
\end{equation}

The values of the atomic parameters \begin{math} \alpha_0 \end{math} and 
\begin{math} \omega_a  \end{math} have been previously tabulated\cite{25}. 
and are listed in Table I. The values of 
\begin{math} g_0  \end{math} and \begin{math} \Omega_s  \end{math},  
(for several metals) have also been previously tabulated\cite{25}  
and are here listed in Table II. In Table III, 
we have computed the values of  \begin{math} C_{atom} \end{math} 
and \begin{math} v_{atom} \end{math}. 

The probability of atomic 
excitation in the Casimir regime    
\begin{math} P_C=(v_{atom}/h^3) \end{math} 
can be calculated in the limit of large 
\begin{math} h  \end{math}. 
When the height of the atom obeys 
\begin{math} h^3>>v_{atom} \end{math}, 
the perturbation theory is reliable. However, 
for atoms adsorbed on a submonolayer of film the height 
\begin{math} h \end{math} is not large. Thus a non-perturbative 
method is required. We now turn our attention to this 
more detailed treatment.

\begin{table}
\caption{Values of $\alpha_0$ and $\hbar \omega_a$ for Noble gas atoms.}
\label{Table1}
\begin{tabular}{ccc}
Atom & ($\alpha_0/\AA^3$) & ($\hbar \omega_a/eV$) \\
\hline 
He  & 0.205 & 27.645 \\ 
Ne & 0.396 & 32.734 \\
Ar & 1.642 & 18.971 \\
Kr & 2.487  & 16.478 \\
Xe & 4.018 & 14.34 
\end{tabular}
\end{table}

\begin{table}
\caption{Values of $g_0$ and $\hbar \Omega_s$ for some metals.}
\label{Table2}
\begin{tabular}{ccc}
Atom & ($g_0$) & ($\hbar \Omega_s/eV$) \\
\hline 
Cu & 0.857 & 17.74 \\ 
Ag & 0.812 & 21.768 \\
Au & 0.840 & 24.162 \\
Al & 0.976 & 12.87 \\
Pd & 0.785 & 17.115 \\
Gr & 0.619 & 18.149 
\end{tabular}
\end{table}

\begin{table}
\caption{$C_{atom}$ and $v_{atom}$ for noble 
gas atoms.}
\label{Table3}
\begin{tabular}{ccc}
Gas-Metal & ($C_{atom} /\AA^3 eV$) & ($v_{atom} /\AA^3$) \\
\hline
Xe-Cu & 3.413 & 0.106 \\ 
Xe-Ag & 3.525 & 0.098 \\
Xe-Au & 3.797 & 0.099 \\
\hline
Kr-Cu & 2.276 & 0.067 \\ 
Kr-Ag & 2.367 & 0.062 \\
Kr-Au & 2.558 & 0.063 \\
\hline
Ar-Cu & 1.612 & 0.044 \\ 
Ar-Ag & 1.690 & 0.041 \\
Ar-Au & 1.832 & 0.043 \\
\hline
Ne-Cu & 0.488 & 0.096 \\ 
Ne-Ag & 0.525 & 0.096 \\
Ne-Au & 0.578 & 0.010 \\
\hline
He-Cu & 0.237 & 0.005 \\ 
He-Ag & 0.253 & 0.005 \\
He-Au & 0.277 & 0.005  
\end{tabular}
\end{table}

\section{Rigorous Results}

The above considerations are true for atoms 
above a metal in the perturbative limit 
\begin{math} h\to \infty  \end{math}. 
When the atom is adsorbed on the metal surface, a non-perturbative 
viewpoint must be invoked\cite{26,27}. For example, consider the 
ground state matrix element analytic in the upper half complex energy 
plane 
\begin{math} {\Im} m\ z > 0  \end{math},
\begin{equation}
G(z)=\left(\psi_0,\left\{
{1\over z-{\cal H}}
\right\}\psi_0\right),
\end{equation}
where \begin{math} \psi_0 \end{math} is the unperturbed ground state,  
\begin{equation}
H\psi_0 =E_0\psi_0,
\end{equation}
and the binding energy of the adsorbed atom, 
\begin{equation}
U=-B,\ \ B>0,
\end{equation}
is determined by the full electronic ground state 
\begin{equation}
{\cal H}\Psi_0=(H+V)\Psi_0=(E_0+U)\Psi_0.
\end{equation}

One may write Eq.(43) in the exact form
\begin{equation}
G(z)=\left\{ {1\over z-E_0-\Sigma (z)}\right\},
\end{equation}
where the self energy part \begin{math} \Sigma (z) \end{math} 
determines the atomic binding energy 
\begin{equation}
U=\Sigma (E_0+U).
\end{equation}
\begin{math} G(z) \end{math} has a simple pole at the 
exact ground state energy \begin{math} z_0=E_0+U=E_0-B \end{math}. 
The residue at the pole is the transition probability 
\begin{math} \psi_0\to \Psi_0  \end{math}; i.e. 
\begin{equation}
|(\Psi_0,\psi_0)|^2=\left({1\over 1-\Sigma^\prime (E_0+U)}\right),
\end{equation}
where \begin{math} \Sigma^\prime (z)=d \Sigma (z)/dz \end{math}.
The exact expression for the self energy part reads 
\begin{equation}
\Sigma (z)=\left(\psi_0,V\left\{
{1\over z-{\cal H}^\prime }
\right\}V\psi_0\right),
\end{equation}
where \begin{math} {\cal H}^\prime =\hat{P}{\cal H} \hat{P} \end{math} 
and where \begin{math} \hat{P}  \end{math} projects into the subspace 
normal to the unperturbed ground state. For example, from Eq.(49) and 
the definition of  \begin{math} \hat{P} \end{math} one finds the 
mean value \begin{math} P=(\Psi_0, \hat{P} \Psi_0) \end{math} 
given by 
\begin{equation}
P=1-|(\Psi_0,\psi_0)|^2=
\left({\Sigma^\prime (E_0+U)\over 1-\Sigma^\prime (E_0+U)}\right).
\end{equation}
The details of the mathematical derivation of Eqs.(47)-(51) are 
given in the appendix.

Suppose that the atom (located on the metal surface) starts out 
in the unperturbed state \begin{math} \psi_0 \end{math}. The atom 
will then decay into its renormalized ground state, giving up its 
excess binding energy \begin{math} U  \end{math} to the bulk 
electrons in the metal. The 
transition rate per unit time to deposit an electronic energy 
\begin{math} W  \end{math} to the metal is given by  
\begin{equation}
\Gamma (W)=\left({2\pi \over \hbar }\right)
\big(\psi_0,V\delta(E_0+W- {\cal H}^\prime )V\psi_0\big).
\end{equation}
From Eqs.(50) and (52) it follows that 
\begin{equation}
\Sigma (z)=\left({\hbar \over 2\pi }\right)
\int_0^\infty {\Gamma (W)dW \over z-(E_0+W)}\ .
\end{equation} 
Thus the binding energy \begin{math}U=-B\end{math}
is rigorously determined by 
Eqs.(48) and (53) to be 
\begin{equation}
B=\left({\hbar \over 2\pi }\right)
\int_0^\infty {\Gamma (W)dW \over W+B}\ ,
\end{equation}  
while Eqs.(51) and (53) imply 
\begin{equation}
P=\left({\varpi \over 1+\varpi}\right),
\end{equation} 
where 
\begin{equation}
\varpi =\left({\hbar \over 2\pi }\right)
\int_0^\infty {\Gamma (W)dW \over (W+B)^2}\ ,
\end{equation} 

In the perturbative limit, 
\begin{equation}
\lim_{h\to \infty} (P/P_C)=\lim_{h\to \infty} (U/U_C)=1
\end{equation}
where the Casimir values \begin{math} U_C \end{math} and 
\begin{math} P_C \end{math} have been defined, respectively, 
in Eqs.(35) and (37). For the non-perturbative limit of atomic 
adsorption for a finite height \begin{math}h=h_s\end{math}, 
the exact results of Eqs.(54), (55) and (56) can be employed for making 
realistic estimates of the effective charge 
\begin{math} Z_{eff}|e|\end{math} of the atom.

\section{Net Charge on an Adsorbed Atom}

For estimating the effective charge on the adsorbed atom, we 
note the following: (i) Eq.(54) is an implicit equation for 
the binding energy \begin{math} B=f(B) \end{math}. (ii) If 
\begin{math} \bar{\Gamma }=\lim_{W\to 0}\Gamma (W)>0 \end{math}, then 
\begin{math} 
f(B\to 0)\approx (\hbar \bar{\Gamma } /2\pi )ln(const/B) 
\end{math}, 
where a constant ``cut-off'' must be placed in the logarithm. 
(iii) We employ the following simple dispersion formula for    
\begin{math} \Gamma(W) \end{math}: Over a bandwidth 
\begin{math} 0<W<E_a \end{math}, we consider 
\begin{math} \Gamma(W)=\bar{\Gamma} \end{math} to be uniform.
Outside this interval, i.e. for electronic energies above the atomic 
energy cut-off \begin{math} E_a=\hbar \omega_a \end{math} in Table I, 
we consider \begin{math} \Gamma(W) \end{math} to be negligible. Under this 
assumption, the binding energy Eq.(54) reads 
\begin{equation}
B\approx\left({\hbar \bar{\Gamma} \over 2\pi }\right)
ln\left({\hbar \omega_a \over B }\right),
\end{equation}
if \begin{math} B<<\hbar \omega_a  \end{math}. Similarly, Eq.(56) 
reads 
\begin{equation}
\varpi \approx \left({\hbar \bar{\Gamma} \over 2\pi B}\right),
\end{equation}

The probability of excitation \begin{math} P=Z_{eff} \end{math} 
determines the effective charge via Eqs.(55), (58) and (59) 
\begin{equation}
Z_{eff}=\left({1\over 1+ln(E_a/B)}\right).
\end{equation}
A plot of the effective charge versus the binding energy is shown in Fig.2.

\begin{figure}[htbp]
\begin{center}
\mbox{\epsfig{file=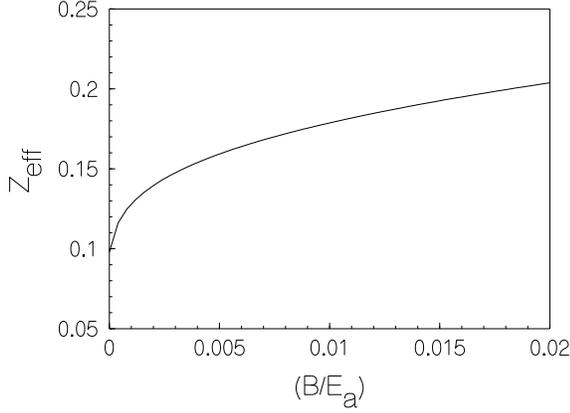,height=80mm}}
\caption{The effective charge $Z_{eff}|e|$ on the atom is shown as a 
function of the binding energy $B$ in units of the atomic 
excitation energy $E_a=\hbar \omega_a$. See Table I.}
\label{fig2}
\end{center}
\end{figure}

In terms of experimental binding energies and mean atomic excitation 
energies listed Table I, the effective charges can be computed numerically 
from Eq.(60). 

\section{Conclusions}

A dispersion theory has been presented for computing the effective 
charge per noble gas atom adsorbed on metallic substrates. The general 
physical situation, shown in Fig.1, is complimentary to the charge 
density functional approach for computing atomic binding.\cite{28,29} 
Both approaches yield a picture in which there is an atomic dipole 
moment. The negative end of the dipole moment is better viewed as the 
donation of a mean negative electronic charge to the metal below. The 
positive end of the atomic dipole moment, is what has been called the 
net atomic charge \begin{math} Z_{eff}|e| \end{math}. 

The advantage to the dispersion method is that it fits smoothly into 
the power law limit of the well known Casimir force. In the density 
functional approach, the ``image charges'' are added in ``by hand''. 
On the other hand, the charge density functional approach is 
intuitive on a microscopic level.

The evidence for \begin{math} Z_{eff}>0\end{math} is based on the 
change in metallic work function due to adding submonolayer adatoms, 
for thicker films the situation is complex.\cite{17} The work function 
can {\it increase} as the film thickness increases for film layers 
above the monolayer. This experimental fact has (thus far) no simple 
theoretical explanation. 
\par \noindent 
\vskip 0.2 cm
\centerline{\bf APPENDIX}
\medskip

In this appendix, the details of the mathematical derivation of results 
in Sec. V will be made explicit. One may define two projection operators 
$\hat{Q}$ and $\hat{P}$ obeying 
$$
\hat{Q}+\hat{P}=1. \eqno(A1)
$$
The operator $\hat{Q}$ projects a wave function onto the 
unperturbed ground state wave function $\psi_0$ and may be written 
as 
$$
\hat{Q}=\left|\psi_0 \right>\left<\psi_0 \right|. \eqno(A2)
$$
If we define for any operator $\hat{A}$, the projected operators 
$A_{QQ}=\hat{Q}\hat{A}\hat{Q}$, 
$A_{QP}=\hat{Q}\hat{A}\hat{P}$, $A_{PQ}=\hat{P}\hat{A}\hat{Q}$ and 
$A_{PP}=\hat{P}\hat{A}\hat{P}$, then the operator may be written in a 
partitioned matrix form 
$$
\hat{A}=\pmatrix{A_{QQ} & A_{QP}\cr A_{PQ} & A_{PP}}. \eqno(A3)
$$
In particular, the resolvent operator 
$$
\hat{\cal R}(z)=\left({1\over z-{\cal H}}\right) \eqno(A4)
$$
may be written as 
$$
\hat{\cal R}(z)=\pmatrix{{\cal R}_{QQ}(z) & {\cal R}_{QP}(z) 
\cr {\cal R}_{PQ}(z) & {\cal R}_{PP}(z)}, \eqno(A5)
$$
where 
$$
G(z)1_{QQ}={\cal R}_{QQ}(z)=
1_{QQ}\sum_n {|(\Psi_n ,\psi_0)|^2 \over (z-{\cal E}_n) }. 
\eqno(A6)
$$
From Eq.(A6), it follows that $G(z)$ has a ground state pole 
at $z_0={\cal E}_0$ with a residue given by $|(\Psi_n ,\psi_0)|^2$; 
i.e. 
$$
G(z)\to {|(\Psi_0 ,\psi_0)|^2 \over (z-z_0)}\ \ \ {\rm as}
\ \ \ z\to z_0 . \eqno(A7)
$$ 
Eqs.(A4) and (A5) imply 
$$
\pmatrix{z-{\cal H}_{QQ} & -{\cal H}_{QP}
\cr -{\cal H}_{PQ} & z-{\cal H}_{PP}}
\pmatrix{{\cal R}_{QQ}(z) & {\cal R}_{QP}(z) 
\cr {\cal R}_{PQ}(z) & {\cal R}_{PP}(z)}
$$
$$
=\pmatrix{1_{QQ} & 0 \cr 0 & 1_{PP}},
\eqno(A8)
$$ 
from which 
$$
(z-{\cal H}_{QQ}){\cal R}_{QQ}(z)-{\cal H}_{QP}{\cal R}_{PQ}(z)
=1_{QQ}, \eqno(A9)
$$
and 
$$
-{\cal H}_{PQ}{\cal R}_{QQ}(z)+(z-{\cal H}_{PP}){\cal R}_{PQ}(z)=0.
\eqno(A10)
$$
Eqs.(A9) and (A10) imply 
$$
\left(z-{\cal H}_{QQ}-{\cal H}_{QP}{1\over z-{\cal H}_{PP}}
{\cal H}_{PQ}\right){\cal R}_{QQ}(z)=1_{QQ}. \eqno(A11)
$$
Finally, if ${\cal H}=H+V$, $(\psi_0,V\psi_0)=0$ and 
${\cal H^\prime }={\cal H}_{PP}$, then Eqs.(A6) and (A11) read 
$$
\left\{z-E_0-\left(\psi_0,V\left\{
{1\over z-{\cal H}^\prime}\right\}V\psi_0\right)
\right\}G(z)=1; \eqno(A12)
$$
i.e. 
$$
G(z)=\left({1\over z-E_0-\Sigma (z)}\right), \eqno(A13)
$$
with a self energy 
$$
\Sigma (z)=\left(\psi_0,V
\left\{{1\over z-{\cal H}^\prime}\right\}V\psi_0\right), \eqno(A14)
$$
which completes our derivations.

\end{document}